\documentclass[10pt,conference]{IEEEtran}
\IEEEoverridecommandlockouts
\usepackage{amsmath,amssymb,amsfonts}
\usepackage{siunitx}
\usepackage{braket}
\usepackage{algorithm2e}
\usepackage{graphicx}
\newcommand\latency{\qty{988.8}{\nano\second} }
\newcommand\pat{hardware-guided pruning and retraining}
\usepackage{ulem} 
\newcommand\lermwpm{{$1.69\times10^{-5}$}}
\newcommand\lergnnsw{{$8.27\times10^{-6}$}}

\newcommand\lergnnpruned{{$1.34\times10^{-5}$}}

\newcommand\lergnnprunedquantizednodes{{$1.47\times10^{-5}$}}
\newcommand\imprgnnprunedquantizednodes{{$13\%$}}
\newcommand{\lerparallel}{{$1.01\times10^{-5}$}}
\newcommand{\imprparallel}{{$40\%$}}
\newcommand{\singlegraph}{\texttt{max-latency}}
\newcommand{\multgraph}{\texttt{average-latency}}

\makeatletter
\newcommand{\linebreakand}{%
  \end{@IEEEauthorhalign}
  \hfill\mbox{}\par
  \mbox{}\hfill
  \begin{@IEEEauthorhalign}
}
\makeatother
\def\BibTeX{{\rm B\kern-.05em{\sc i\kern-.025em b}\kern-.08em
    T\kern-.1667em\lower.7ex\hbox{E}\kern-.125emX}}
\begin{document}

\title{Low-Latency GNN Accelerator for\\ Quantum Error Correction}

\author{
\IEEEauthorblockN{Alessio Cicero}
\IEEEauthorblockA{\textit{Chalmers University of Technology}\\
\textit{and University of Gothenburg}\\
Gothenburg, Sweden\\
ciceroa@chalmers.se}
\and
\IEEEauthorblockN{Luigi Altamura}
\IEEEauthorblockA{\textit{Chalmers University of Technology}\\
\textit{and University of Gothenburg}\\
Gothenburg, Sweden\\
altamura@chalmers.se}
\and
\IEEEauthorblockN{Moritz Lange}
\IEEEauthorblockA{\textit{University of Gothenburg}\\
Gothenburg, Sweden\\
moritz.lange@physics.gu.se}
\linebreakand
\IEEEauthorblockN{Mats Granath}
\IEEEauthorblockA{\textit{University of Gothenburg}\\
Gothenburg, Sweden\\
mats.granath@physics.gu.se}
\and
\IEEEauthorblockN{Pedro Trancoso}
\IEEEauthorblockA{\textit{Chalmers University of Technology}\\
\textit{and University of Gothenburg}\\
Gothenburg, Sweden\\
ppedro@chalmers.se}
}

\maketitle

\begin{abstract}
Quantum computers can solve select classes of problems in a much more efficient way than classical computers, but current implementations are limited by high physical error rates. Quantum Error Correction~(QEC) codes address this issue by encoding multiple physical qubits into a logical qubit to achieve a lower logical error rate, with the surface code being one of the most commonly used.\par
Because syndrome measurements are produced continuously during operation, the decoder must process them within strict time constraints to avoid becoming a system bottleneck. As a result, real-time decoding is a fundamental requirement for fault-tolerant quantum computing. While most of the state-of-the-art real-time decoders are based on Minimum-Weight Perfect Matching~(MWPM), due to its strong trade-off between decoding accuracy and practical implementability, in this work we choose a high-accuracy Graph Neural Network~(GNN) that trades higher computational complexity for lower logical error rates. \par
To make this GNN practical for real-time decoding, we adopt an algorithm-hardware co-design approach. We first reduce its complexity through hardware-guided pruning and retraining, obtaining two hardware-friendly models that reduce parameter count by $3.1\times$ and $6.5\times$, targeting an average decoding latency of one syndrome cycle and a worst-case latency within one syndrome cycle, respectively. We further reduce the hardware cost through input-graph filtering and post-training quantization. Finally, we propose an FPGA-based architecture designed around these two pruned, quantized, and graph-bounded models and optimized for low-latency inference, enabling real-time decoding.\par
Evaluated on surface codes up to distance 7, circuit-level noise model, and physical error rate p=$10^{-3}$ our decoder shows clear advantages over MWPM in decoding accuracy in both \multgraph{} and \singlegraph{} decoding, reducing logical error rate by 40\% at \qty{1}{\micro\second} average latency in the former and by 13\% under a strict \qty{1}{\micro\second} deadline in the latter.
\end{abstract}


\section{Introduction}%1pg
\label{sec:intro}

Quantum computers have the potential to revolutionize several domains by offering computational speed-ups compared to the execution on classical machines. Notable examples are complex problems such as developing new chemical compounds~\cite{quantum-advantage-chemistry,motta2022emerging-quantum-chemistry1,lee2023evaluating-quantum-chemistry2} or evaluating the physical properties of new materials~\cite{quantum-advantage-material,alexeev2024quantum-quantum-material}.
However, today's quantum devices remain fundamentally limited by the high error rate of their physical qubits~\cite{battistel2023real-error-limiting-factor}, the basic unit of a quantum computer~\cite{Nielsen_Chuang_2023}. 
% These errors, which may take the form of bit flips or phase flips~\cite{Nielsen_Chuang_2023}.

A useful approximation of the physical error rate is the ratio between gate execution time and qubit lifetime. Lower error rates allow running longer sequences of reliable quantum operations. In current superconducting hardware, qubit lifetimes are on the order of a few hundred microseconds and gate times on the order of a hundred nanoseconds, giving error rates on the order of $10^{-3}$. Developing an effective implementation of Quantum Error Correction~(QEC)~\cite{chatterjee2023quantum-qecfordummies} is therefore necessary to support more complex algorithms, which may require millions or billions of gates, and move towards the era of Fault-Tolerant Quantum Computing~(FTQC)~\cite{Nielsen_Chuang_2023}. 

\begin{figure}
    \centering
    \includegraphics[width=1.0\linewidth]{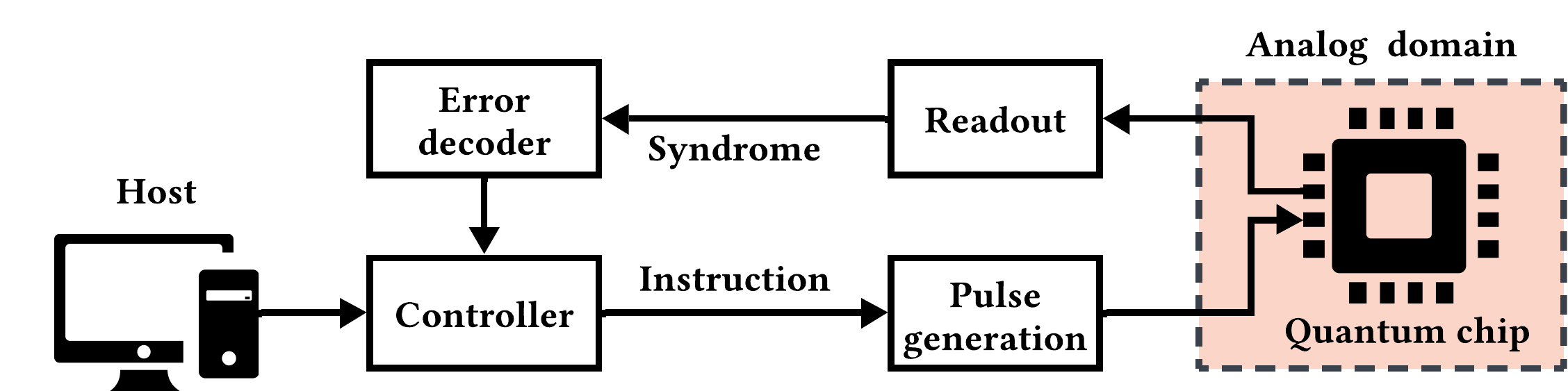}

    \caption{The host computer sends a quantum program to the controller, which drives the quantum chip with precise control pulses. The outputs of the chip are measured, and the error syndromes are computed from the observed errors. The controller may apply corrections based on those results.}
    \label{fig:error-correction-cycle}
\end{figure}
\par

Among the possible QEC codes proposed in the literature for superconducting qubits, the surface code~\cite{fowler2012surface} has emerged as one of the most promising approaches. This is due to its high threshold, that is, the physical error rate below which increasing code size reduces the logical error rate, and its compatibility with two-dimensional superconducting qubit layouts~\cite{google2023suppressing-surfacecode}. By encoding multiple physical qubits as a single logical qubit, the surface code achieves an overall logical error rate lower than the error rate of the individual physical qubits. Its error correction capability is connected to the code distance~(d), which defines both the code size and number of physical qubits involved~\cite{fowler2012surface}.\par 
In the surface code, errors are detected through repeated syndrome measurements that are used to construct a syndrome graph. Decoding - the process of identifying the most likely error chains from this graph - is the most time-critical operation in QEC, since each graph must be processed quickly enough to avoid incurring a backlog of measurements~\cite{terhal2015quantum}. The overall execution flow is shown in Figure~\ref{fig:error-correction-cycle}. In the case of the surface code for superconducting qubits~\cite{superconducting}, the syndrome measurement rounds can be as fast as \qty{1}{\micro\second} ~\cite{google2023suppressing-surfacecode}. Two different timing constraints have been considered in prior work: either decoding within the latency of a single syndrome measurement round~(\singlegraph{})~\cite{alavisamani2024promatch,das2022lilliput}, or sustaining an average decoding rate faster than syndrome extraction~(\multgraph{})~\cite{terhal2015quantum,skoric2023parallel-parallel-decoding-nature, wu2025micro-state-of-art-decoder}. In this work, we will consider both.
% the decoder must operate at the pace of syndrome extraction, which can be as fast as \qty{1}{\micro\second} per QEC round~\cite{google2023suppressing-surfacecode}. In \singlegraph{} decoding, where a correction is applied after each measurement round, this rate sets the maximum tolerable decoding latency~\cite{skoric2023parallel-parallel-decoding-nature}. By contrast, in \multgraph{} decoding~\cite{skoric2023parallel-parallel-decoding-nature}, a recently proposed scheme, the timing constraint is more relaxed: the decoder needs only to sustain, on average, the throughput of one measurement round.

% This introduces a heavy constraint on the available decoding time, which must be respected to have real-time quantum error correction. \par
%Classical Decoding challenges
A range of decoding algorithms exists for the surface code, trading off latency against logical error rate. Most hardware accelerators for real-time decoding are based on MWPM~\cite{wu2025micro-state-of-art-decoder, alavisamani2024promatch, vittal2023astrea}, due to its balance between decoding accuracy and implementation cost, which allows to achieve real-time decoding up to $d=13$~\cite{wu2025micro-state-of-art-decoder}. In our work, we instead use a higher accuracy decoder based on a graph neural network~(GNN)~\cite{lange2023data}, which has been shown to outperform MWPM in terms of logical error rate for code distances up to $7$. This matches the largest code distance demonstrated in current physical experiments~\cite{acharya2024quantum-google-chip}. 
% Moreover, the original results of~\cite{lange2023data} indicate that further scaling to higher code distances is achievable with larger GNN models. Such models could be supported either by future FPGA platforms with greater available resources or by moving to decoding schemes with relaxed latency constraints, such as sliding-window decoding~\cite{google2025quantum-sliding-window} or parallel decoding~\cite{skoric2023parallel-parallel-decoding-nature}.
\par 
In the context of real-time decoding for QEC, software-based decoders are too slow to meet the latency constraint of \qty{1}{\micro\second}, in both \singlegraph{} decoding and \multgraph{} decoding schemes. The most common approach to real-time decoding, as also shown in the previously mentioned works~\cite{wu2025micro-state-of-art-decoder,alavisamani2024promatch,vittal2023astrea}, is to design an FPGA-based accelerator optimized for low-latency decoding. \par
Similarly, this GNN~\cite{lange2023data} poses a significant challenge for real-time decoding due to its large model size, with approximately $10^6$ parameters, and a number of computations that grow polynomially with the input graph size. To make this GNN-based decoder viable for real-time decoding, we must address both its high complexity and its polynomial scaling at the system level, and co-design the architecture to ensure that the final system satisfies the target latency.\par
Starting from the GNN described in~\cite{lange2023data}, we first characterize the main latency bottlenecks of the decoder: the high number of multiplications and their scaling with the number of nodes of the input graph. This characterization leads to an algorithm-hardware co-design methodology. \par
At the algorithm level, we analyze the trade-off introduced by reducing the number of layers using \pat{}, where improved latency is obtained by partially reducing the logical error rate advantage over MWPM. Based on this analysis, we derive two different models, one aggressively pruned to target \singlegraph{} decoding latencies, and one pruned more conservatively with a lower logical error rate, for \multgraph{} decoding. Respectively, the two models have a $6.5\times$ and $3.1\times$ reduction in the number of parameters compared to the unpruned GNN. We additionally bound the input-graph size through input-graph filtering, limiting the supported graph size by~$\approx80\%$ with limited logical error rate increase.
\par
On the hardware side, we design a low-latency hardware architecture that supports both models with only minor adaptations by optimizing the resource allocation, execution schedule, and weight delivery to the pruned layer structure. 
\par
We achieve real-time decoding while reducing the logical error rate compared to the state-of-the-art in both settings. The larger model improves over MWPM logical error rate by \imprparallel{} in \multgraph{} decoding, achieving a logical error rate of \lerparallel{} while decoding on average under \qty{1}{\micro\second}, while the smaller model improves over MWPM by \imprgnnprunedquantizednodes{} in \singlegraph{} decoding, achieving a logical error rate of \lergnnprunedquantizednodes, while always decoding within \qty{1}{\micro\second}. In both cases, the reduction in logical error rate corresponds to supporting proportionally longer quantum circuits without increasing the physical-qubit budget.

Our key contributions include:
\begin{itemize}
    \item Characterization of the hardware-guided pruning and retraining design space of this GNN-based QEC decoder, targeting a reduction in complexity while limiting the corresponding increase in logical error rate.
    \item Multiple hardware-aware approaches, including post-training quantization, input-graph filtering, and hardware-informed architectural decisions, designed to meet the distinct latency targets of \multgraph{} and \singlegraph{} decoding. 
    \item A custom-designed FPGA accelerator for the GNN-based decoder that outperforms MWPM-based decoders in logical error rate in both \multgraph{} and \singlegraph{} decoding, while running under the low-latency constraints.
    
\end{itemize}

\section{Background}%1pg
\label{sec:background}
% \subsection{Quantum computing}
% \textcolor{blue}{Do we need this paragraph?}
 Superconducting physical qubits currently exhibit high physical error rates (typically \( 10^{-3} \)~\cite{Jurcevic_2021-physical-qubit-error-rate}), far above what is tolerable for any useful applications of quantum algorithms. For example, factoring a 2048-bit RSA integer with Shor's algorithm would require a logical error rate of \( 10^{-15} \)~\cite{gidney2025factor-rsa20468}. QEC achieves much lower logical error rates by encoding information using multiple physical qubits and decoding errors in real time. When combined with fault-tolerant gate implementations~\cite{postler2022demonstration-fault-tolerant-quantum-gates}, QEC allows for increasing the number of sequential quantum computations carried out reliably, even in the presence of noise~\cite{terhal2015quantum}.

\subsection{Quantum Error Correction}

Quantum systems are inherently fragile due to their susceptibility to decoherence and noise from interactions with the environment. 
\par
Unlike classical bits, which can be easily stored and measured, qubits exist in superpositions of states and collapse upon measurement. Moreover, quantum information cannot be copied (as per the no-cloning theorem), which makes traditional error correction techniques unfeasible~\cite{Nielsen_Chuang_2023}. \par

To preserve the integrity of the logical qubit, QEC protocols encode the logical information in a set of data qubits. Additional qubits~(often referred to as ancilla qubits or measure qubits~\cite{Nielsen_Chuang_2023,fowler2012surface}), are dedicated solely to measurement operations. These interact with the data qubits and are then measured to extract error syndromes: the outcomes of measurements that indicate which type of error (if any) occurred, without revealing or disturbing the encoded state. 
To quantify decoding performance, we can define the logical error rate as a function of correctly decoded syndromes over the total number of syndromes:
\begin{equation}
\text{Logical\;error\;rate} = 1 - \frac{\textit{Correctly decoded syndromes}}{\textit{Total\;syndromes}}
\end{equation}

\subsection{Surface Code}

\begin{figure}
    \centering
    \includegraphics[width=0.8\linewidth]{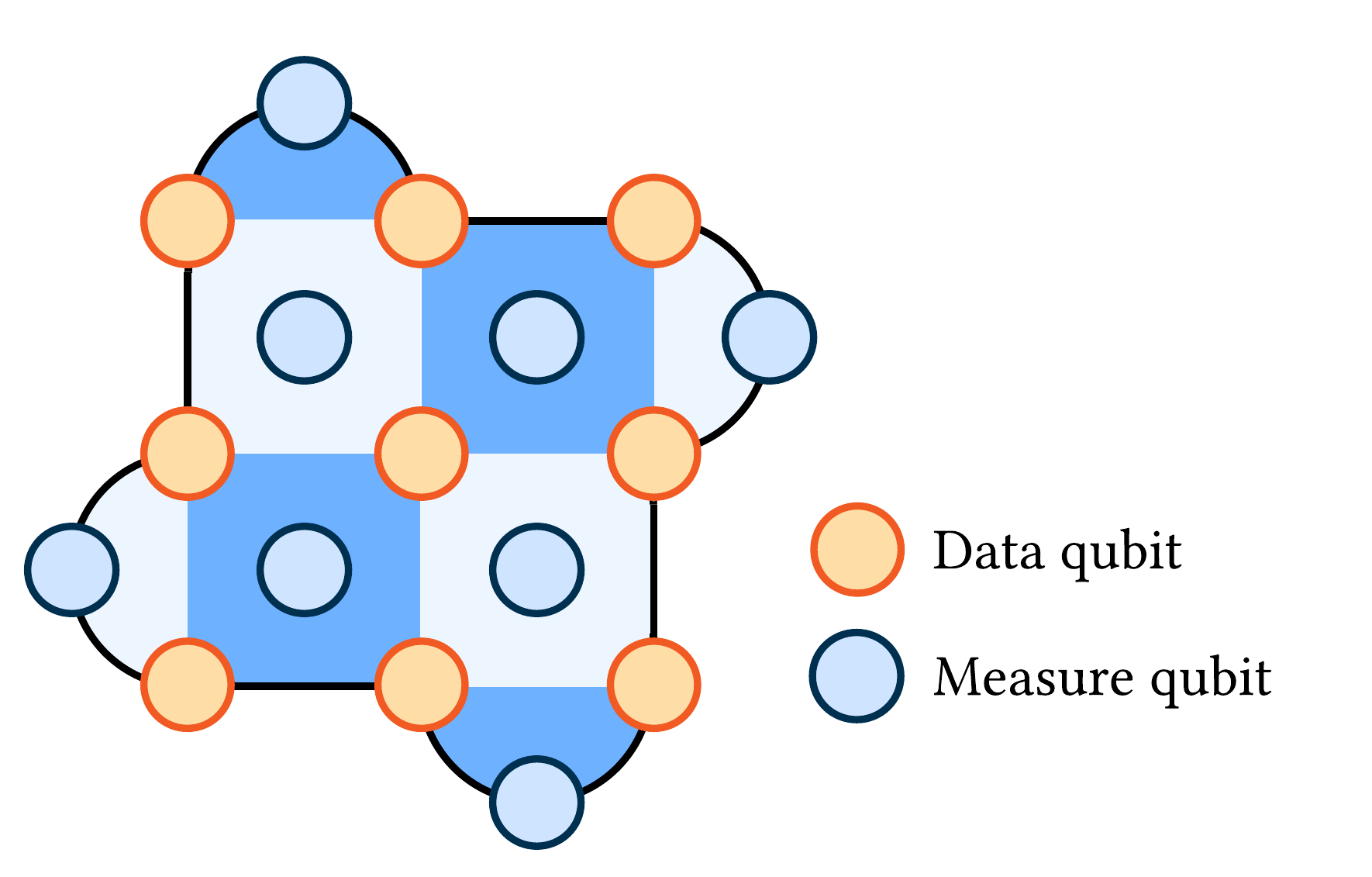}
    \caption{Surface code of distance 3, with qubits highlighted according to their roles as data or measurement qubits.}
    \label{fig:surface-code}
\end{figure}

The surface code is one of the most prominent QEC codes due to its high error threshold, locality of measuring quantum circuits, and suitability for hardware with planar layouts and nearest-neighbor interactions, making it one of the leading candidates for practical fault-tolerant quantum computation. As shown in Figure~\ref{fig:surface-code}, its underlying graph aligns well with the physical qubit connectivity of actual quantum hardware~\cite{google2023suppressing-surfacecode}. 
Figure~\ref{fig:surface-code} illustrates the relationship between code distance~(d) and the number of physical qubits used to encode a single logical qubit. Specifically, $d^2$ data qubits and $d^2-1$ measure qubits are required.  Measurements are performed using dedicated quantum circuits, defining stabilizers~\cite{Nielsen_Chuang_2023}, between the data qubits and the measure qubits. The surface code can correct up to $\lfloor\frac{d-1}{2}\rfloor$ single-qubit errors~\cite{fowler2012surface}.
Lastly, to account for circuit-level noise, $d\_t$ repeated measurements are required to detect data qubit errors and also potential readout errors; in this work, we select the standard  $d\_t=d$. Multiple rounds of measurement help distinguish genuine changes in data qubit states from spurious results caused by noise affecting the measurement qubits or the readout process. \par
As these rounds are generated continuously, decoding must satisfy strict timing requirements. In the case of superconducting qubits, prior work has considered two types of timing constraints to avoid a backlog of errors: decoding within a single syndrome measurement round~(\singlegraph{})~\cite{alavisamani2024promatch,das2022lilliput}, or sustaining an average decoding throughput higher than the syndrome extraction rate~(\multgraph{})~\cite{terhal2015quantum,skoric2023parallel-parallel-decoding-nature,wu2025micro-state-of-art-decoder}. In the first timing model, each syndrome graph must be decoded before the next one is produced, as each syndrome is decoded independently. In the second one, used in settings such as sliding window~\cite{google2025quantum-sliding-window} or parallel decoding~\cite{skoric2023parallel-parallel-decoding-nature}, the system can tolerate latency variation as long as the average decoding rate remains higher than the syndrome generation rate to avoid accumulating a backlog of errors. As both regimes have been the focus of recent work~\cite{wu2025micro-state-of-art-decoder,alavisamani2024promatch}, we target both in this paper.

\subsection{Decoding Algorithms}
% In any quantum error correcting code, physical errors manifest as changes in stabiliser measurement outcomes, known as the \emph{syndrome}. 
In QEC codes, physical errors manifest as changes in syndrome measurements. 
Decoding is the process of interpreting this syndrome to identify the most likely (logical) set of errors that occurred,
% \textcolor{blue}{(more precise definition of the decoding problem?)}
 preserving the encoded logical information.  \par
Efficient and accurate decoding is critical for maintaining fault tolerance, especially under the stringent time constraints imposed by the quantum hardware. 
% \color{red}
% This requirement arises because
% % , as illustrated in Figure~\ref{fig:surface-code}, 
% the decoder must operate within the duration of a single measurement cycle, as also previously discussed in Section~\ref{sec:intro}.
% % \textcolor{blue}{(mention 1 microsecond here?)}. 
% \color{black}
A variety of decoding algorithms have been developed, each balancing trade-offs between accuracy, computational complexity, and suitability for hardware acceleration. The most common approaches are summarized here: 
% \textcolor{blue}{more references?}:
% \textcolor{blue}{Alessios summary below:}
\begin{itemize}
    \item \textbf{Union Find~(UF) decoder}~\cite{delfosse2021almost-unionfind}: A lightweight and low-complexity decoding algorithm, that achieves a higher logical error rate compared to the other more advanced methods.
    \item \textbf{Minimum Weight Perfect Matching~(MWPM) decoder}~\cite{kolmogorov2009blossom}: The most widely adopted decoder for surface codes. It offers improved accuracy over Union-Find but comes with higher computational complexity.
    \item \textbf{Belief Propagation~(BP) decoder}~\cite{old2023generalized-belief-propagation}: A more computationally intensive approach than MWPM, particularly suited for decoding quantum Low-Density-Parity-Check~\cite{breuckmann2021quantum-qldpc} codes, where MWPM is less effective.
    \item \textbf{Neural-Network~(NN) decoder}~\cite{varsamopoulos2020decoding-nn-decoder1, varbanov2025neural-nn-decoder2, bausch2023learning-nn-decoder3, lange2023data}: These decoders offer a flexible design space, with accuracy and complexity varying widely depending on the model architecture and training methodology.
\end{itemize}

% The two most used quantum error correction codes are the surface code~\cite{fowler2012surface} and low-density parity-check codes~\cite{breuckmann2021quantum-qldpc}. Surface code based decoders use a variety of algorithms, including minimum weight perfect matching (MWPM) implemented as Blossom Algorithm~\cite{kolmogorov2009blossom}, union find (UF)~\cite{delfosse2021almost-unionfind}, belief propagation (BP)~\cite{old2023generalized-belief-propagation}, and neural networks~\cite{lange2023data}. \par

% \subsection{Logical error rate}
%  To quantify decoding performance, we can define the logical error rate as the proportion of correctly predicted logical errors out of the total number of syndromes:
% \begin{equation}
% Logical\;error\;rate = \frac{Correct\;predictions}{Total\;syndromes}
% \end{equation}

% \subsection{Graph generation}
% \textcolor{blue}{move this to Sec. III?}
% \subsection{Related Work}
The most commonly used decoder for hardware implementation is MWPM~\cite{alavisamani2024promatch,vittal2023astrea,das2022lilliput}, with the current state-of-the-art being Micro Blossom~\cite{wu2025micro-state-of-art-decoder}. The authors are able to decode surface codes of code distance up to $d=13$ while still being under the \qty{1}{\micro\second} threshold, and fitting the hardware decoder in a single  Xilinx Versal VMK180 FPGA. While they are able to achieve the required latency, the logical error rate of the MWPM approach is still worse than that obtainable from optimized neural-network approaches. \par
Another main line of work is on the UF decoding, which achieves reduced decoding complexity at the cost of increased logical error rates than MWPM. However, its simpler algorithm allows for an even faster execution. The state-of-the-art work implementing the UF decoder on FPGA is~\cite{liyanage2024fpga-ufdecoder51}; they are able to decode surface codes up to $d=51$ while still achieving the real-time latency. 
While prioritizing lower-complexity decoders is important for maintaining low latency at higher code distances, high-accuracy decoders achieve lower logical error rates, with the same number of qubits, supporting more complex quantum circuits in the future. For this reason, in our work we adopt a Neural-Network-based decoder, specifically a Graph Neural Network~(GNN) decoder.

\section{GNN-based Decoder}%1pg
\label{subsec:original-gnn}
% \textcolor{blue}{Description of the GNN itself, with graph of the accuracy compared to the MWPM. Also the latency of the GNN software implementation, with details on what machine it has been run. Maybe need for further characterization on clusters or similar?}
\label{sec:gnn}
The GNN decoder used in our work is based on the architecture proposed by Lange et al.~\cite{lange2023data}, designed to decode surface code syndromes under realistic, circuit-level noise. The decoder treats each multiple rounds of stabilizer measurements as a graph, where nodes represent detection events, and edges connect them based on local proximity. Each node is annotated with a feature vector encoding the stabilizer type (X or Z) and its space-time coordinate.

The network follows a message-passing paradigm composed of the following stages:

\begin{itemize}
\item \textbf{Input Encoding:} Each node starts with a feature vector \( \vec{X}_i^{(0)} \in \mathbb{R}^5 \) representing local information as discussed above.

\item \textbf{Graph Convolution~(GraphConv) Layers:} A sequence of \( L \) message-passing layers propagates information across the graph. Each layer $\ell$ updates node embeddings via:
    \begin{equation}
    \label{eq:gcn}
    \vec{X}_i^{\ell + 1}=\sigma\left(W^\ell_1 \vec{X}^\ell_i+W^\ell_2 \sum_j e_{i j} \vec{X}^\ell_j+\vec{b}^\ell\right)
    \end{equation}
where $W_1^\ell$, $W_2^\ell$ and $\vec{b}^\ell$ are the trainable weights of layer $\ell$ and the element-wise acting rectified linear unit, $\sigma(x)=$ $\operatorname{ReLU}(x)=\max (0, x)$.
\item \textbf{Global Mean-Pooling~(GMP):} After message passing, node embeddings are aggregated using mean-pooling to form a graph-level embedding: 
\begin{equation}
\label{eq:gmp}
\vec{X}_{\text {mean}}=N^{-1} \sum_i \vec{X}^L_i,
\end{equation}
for a graph consisting of $N$ nodes. 
\item \textbf{Classification Head:} A final multilayer perceptron composed of fully connected~(dense) layers maps \( \vec{X}_{\text {mean}} \) to a binary output indicating the presence of a logical error. 
\end{itemize}
% \begin{table}
% \centering
% \begin{tabular}{|c|c|c|}
% \hline
% \textbf{Layer}        & \textbf{$d_{in}$} & \textbf{$d_{out}$} \\
% \hline
% GraphConv\textsubscript{1} & 5   & 32  \\
% GraphConv\textsubscript{2} & 32  & 128 \\
% GraphConv\textsubscript{2} & 128  & 256 \\
% GraphConv\textsubscript{2} & 256  & 512 \\
% GraphConv\textsubscript{2} & 512  & 512 \\
% GraphConv\textsubscript{2} & 512  & 256 \\
% Global mean pool & 256 & 256 \\
% Dense\textsubscript{1}   & 256 & 128 \\
% Dense\textsubscript{1}   & 128 & 64 \\
% Dense\textsubscript{out}   & 64  & 1  \\
% \hline
% \textbf{Total parameters}   &\multicolumn{2}{|c|}{1294720}   \\
% \hline
% \end{tabular}
% \caption{Architecture of the original GNN decoder~\cite{lange2023data}.}
% \label{tab:original-gnn}
% \end{table}

% The GNN details are summarized in Table~\ref{tab:comparison-gnn}.

The GNN operates on an input graph constructed from the measured syndrome data. Each syndrome measurement corresponds to a node in the graph. During graph construction, each node is connected to its $k$-nearest neighbors, forming an undirected graph, with $k=10$. Each edge is assigned a weight, which is symmetric with respect to the direction and is computed as the inverse of the square Euclidean distance between the connected nodes.
The resulting input graph consists of $N$ nodes, where the maximum $N$ depends on the code distance. Each node in the graph is characterized by the following components:
\begin{itemize}
\item \textbf{Node features}: Each node is associated with a feature vector of dimension $5$, as the original GNN.
\item \textbf{Edge indices}: A list that encodes graph connectivity by specifying source and target node pairs.
\item \textbf{Edge weights}: A set of scalar weights for each edge, derived from the inverse square of the distance between nodes.
\end{itemize}

% \subsection{Software-based GNN limitations}
Although the software GNN decoder has a lower logical error rate than MWPM, as shown in Table~\ref{tab:accuracy-and-latency-comparison}, its single-graph inference time is one order of magnitude greater - exceeding the latency requirements for real-time surface code decoding.
% \begin{table}[]
% \centering
% \begin{tabular}{|c|c|c|c|}
% \hline
% \textbf{Decoder}        & \textbf{Accuracy} & \textbf{Logical error rate}  & \textbf{Inference time} \\
% \hline
% MWPM~\cite{higgott2022pymatching} &  0.99929 & $7.10\times10^{-4}$&  \textcolor{blue}{*}\\
% GNN~\cite{lange2023data} &  0.99942 & $5.80\times10^{-4}$& \textcolor{blue}{*}\\
% Our GNN &  0.99941 & $5.90\times10^{-4}$ &\textcolor{blue}{*}\\
% \hline
% % \multicolumn{3}{l}{$^{\mathrm{*}}$\textcolor{blue}{will run this as soon as the desktop is free}}
% \end{tabular}
% \caption{Accuracy comparison between the two architectures and MWPM~\cite{higgott2022pymatching}}
% \label{tab:accuracy-and-latency-comparison}
% \end{table}

To achieve the low-latency requirements of real-time decoding, we transition to a custom hardware accelerator, similarly to prior work in the field~\cite{wu2025micro-state-of-art-decoder,vittal2023astrea,liyanage2024fpga-ufdecoder51}. Specifically, we adopt an FPGA-based implementation, which represents the state-of-the-art approach for this class of problems. This choice is further motivated by the fact that other critical components of the quantum computer stack, such as the quantum controller, are also typically implemented on FPGAs~\cite{xu2021qubic-fpga-quantum-controller}, enabling tighter integration and more efficient co-design. \par

\begin{table}
\caption{Logical error rate and single batch inference time per cycle. Comparison between the GNN~\cite{lange2023data} and MWPM~\cite{higgott2022pymatching} software decoders. Experiment conducted on a system equipped with an Intel i9-12900K CPU and an NVIDIA RTX A4000 GPU.}
\centering
\begin{tabular}{ccc}
\hline
\textbf{Decoder}   & \textbf{Logical error rate}  & \textbf{Average inference time} \\
\hline
MWPM~\cite{higgott2022pymatching} & \lermwpm &  \qty{30.3}{\micro\second}\\
GNN~\cite{lange2023data} &  \lergnnsw& \qty{291.4}{\micro\second}\\
\hline
% \multicolumn{3}{l}{$^{\mathrm{*}}$\textcolor{blue}{will run this as soon as the desktop is free}}
\end{tabular}

\label{tab:accuracy-and-latency-comparison}
\end{table}
% \section{}

\section{GNN Hardware Implementation}%2pg
\label{sec:hardware-implementation}

% The GNN has to decode an input graph which is described as follows. Each measured syndrome will generate a node of the graph used as input for the GNN itself. When building the graph, each node is considered connected up to the knn neighbour. Each graph connection is undirected, and there is a weight for each edge, which is the same in both directions. Therefore the input of our graph will be a graph with n nodes, with n function of the code distance, and each node will have the following parameters:
% \begin{itemize}
%     \item Features: In our case we build the graph with 5 input features for each node
%     \item Edge indexes: List of edges which allows to keep track of what node is connected to what
%     \item Edge weights: the weight of the connected node is function of the inverse of the squared distance
% \end{itemize}
% In particular we can plot the statistics of the number of total nodes, to understand what is the maximum number of nodes we can have based on the selected code distance, as shown in Figure~\ref{fig:node-distribuition}.
% \begin{figure}
%     \centering
%     \includegraphics[width=\linewidth]{img/n_nodes_distribution_d_3.pdf}
%     \caption{Node distribution for code distance d=3}
%     \label{fig:node-distribuition}
% \end{figure}
% In the case of code distance 3, we have that we need to compute graphs up to 10 nodes.
% As shown in Section~\ref{sec:gnn}, the software implementation of the GNN has a latency in the order of hundreds of microseconds even when executed on workstations, which is orders of magnitude slower than the required threshold. 
In this section, we present our hardware-aware co-design methodology, aimed at reducing model complexity to fit within the available FPGA resources, while simultaneously reducing overall latency to meet the tight timing constraints. \par
\subsection{Bottleneck Characterization}
The implementation of the previously described GNN in hardware is unfeasible in any currently available FPGA, due to the resource and latency constraints. As presented in Table~\ref{tab:comparison-gnn}, the number of parameters is in the order of millions, which presents a challenge in terms of computation latency. 
% Because we target FPGAs, our resources are limited, even for high-end devices. 
Additionally, as shown, the number of multiplications scales linearly with the number of nodes of the measured surface code graph. \par
This greatly increases the overall number of operations per layer, due to the fact that the maximum number of nodes~($N_{max}$) in the input graph scales cubically with the code distance~($d$). This is because the number of measurements~($d\_t$) is also equal to the code distance, such that:
\[N_{max}=\frac{d^2-1}{2}\cdot d\_t, \quad \quad \text{with }d\_t=d\]
Respectively, for $d=3$, $5$, and 7 the maximum number of nodes is $12$, 60, and 168. 

Assuming the worst-case for $d=7$, we can have up to 168 input nodes, leading to a number of multiplications in the order of $10^8$. 
Considering an average FPGA $f_{clk}$=\qty{300}{\mega\hertz}, and the required maximum or average latency of \qty{1}{\micro \second}, we would need to process $\approx3\cdot10^5$ operations per cycle. As this is too high for the capabilities of current FPGAs, it is necessary to optimize the implementation across multiple levels. \par
Since the two main challenges are the high number of multiplications and their scaling with the size of the input graph node, we start our approach from optimizing the former through pruning and the latter through input graph filtering.
% \begin{table}

% \caption{Comparison between the parameters of the two GNNs~\cite{lange2023data}.}
% \centering
% \begin{tabular}{ccccc}
% \hline
% &\multicolumn{2}{c}{\textbf{GNN}~\cite{lange2023data}}&\multicolumn{2}{c}{\textbf{Our GNN}}\\  
% \hline
% \textbf{Layer}        & \textbf{$d_{in}$} & \textbf{$d_{out}$} & \textbf{$d_{in}$} & \textbf{$d_{out}$} \\
% \hline
% GraphConv\textsubscript{0} & 5   & 32 & 5   & 32 \\
% GraphConv\textsubscript{1} & 32  & 128 & 32   & 128\\
% GraphConv\textsubscript{2} & 128  & 256 & 128   & 128\textsuperscript{*}\\
% GraphConv\textsubscript{3} & 256  & 512 & -   & -\\
% GraphConv\textsubscript{4} & 512  & 512 & -   & -\\
% GraphConv\textsubscript{5} & 512  & 256 & -   & -\\
% GraphConv\textsubscript{6} & 512  & 256 & -   & -\\
% Global mean pool & 256 & 256 & 256   & 256\\
% Dense\textsubscript{0}   & 256 & 256 & 256 & 256\\
% Dense\textsubscript{1}   & 256 & 128 & 256 & 128\\
% Dense\textsubscript{2}   & 128 & 64 & 128 & 32\\
% Dense\textsubscript{out}   & 64  & 1  & 32 & 1\\
% \hline
% \textbf{Total parameters}   &\multicolumn{2}{c}{1360512} &\multicolumn{2}{c}{148096}   \\
% \hline
% \end{tabular}

\begin{table}
\caption{ Number of multiplications required for each layer
of the GNN of the work~\cite{lange2023data}. GraphConv layers are reported as a function of the number of nodes $N$ of the input graph.}
\centering
\begin{tabular}{cccc}
\hline
\textbf{Layer}        & \textbf{$d_{in}$} & \textbf{$d_{out}$} & \textbf{Multiplications} \\
\hline
GraphConv0 & 5   & 32 & $320\times N$\\
GraphConv1 & 32  & 128 & $8{,}192\times N$\\
GraphConv2 & 128  & 256 & $65{,}536\times N$\\
GraphConv3 & 256  & 512 & $262{,}144\times N$\\
GraphConv4 & 512  & 512 & $524{,}288\times N$\\
GraphConv5 & 512  & 256 & $262{,}144\times N$\\
GraphConv6 & 512  & 256 & $131{,}072\times N$\\
GMP & 256 & 256 & $256$\\
Dense0   & 256 & 256 & $65{,}536$\\
Dense1   & 256 & 128 & $32{,}768$\\
Dense2   & 128 & 64 & $8{,}192$\\
DenseOut   & 64  & 1  & $64$\\
% \hline
% \textbf{Total parameters}   &\multicolumn{2}{c}{1360512} &\multicolumn{2}{c}{148096}   \\
\hline
\end{tabular}
\label{tab:comparison-gnn}
\end{table}

\subsection{Hardware-Guided Pruning and Retraining}
\label{subsec:gnn-new}
Led by the necessity to mitigate the high multiplication count, we did a first evaluation of the distribution of the values of each layer output feature vector, obtained by running the GNN inference on $10^8$ input graphs. This highlighted that many output features had a high probability of being zero, due to the ReLU nonlinearities. 
This sparsity is non-uniform both across feature-vector elements, with some elements much more likely to be zero than others, and across layers, with some layers producing a larger number of zero-valued output elements than others. These statistics were consistent across all input graphs, indicating that they are primarily determined by the weights rather than by the specific input. In Table~\ref{tab:sparsity-per-layer} we report for each layer the number of the output vector feature elements that are zero with at least an $80\%$ probability, which we define as the activation sparsity probability.\par
We use this profiling information to derive a hardware-guided pruning metric. We determine the pruning order by ranking layers based on their number of avoidable multiplications, defined as the product of the activation sparsity probability and the layer’s total multiplication count. Reducing the number of layers, and thus the number of parameters, achieves significant reductions in latency and storage, at the cost of a possible increase in the logical error rate.
Layers with the largest number of avoidable multiplications are therefore pruned first, since removing them provides the largest expected reduction in latency and storage. This is particularly beneficial with this GNN because the layers with the highest hardware cost also present substantial activation sparsity, making them strong candidates for pruning.
% This insight led us to examine the impact of pruning layers with the highest activation sparsity on the logical error rate, under the hypothesis that pruning these layers would incur only a minimal degradation in error-correction performance. 
\par

Since post-training pruning~\cite{li2025improving-structured-post-training-pruning,ling2024slimgpt-structured-post-training-pruning,an2024fluctuation-structured-post-training-pruning} - removing layers without retraining - resulted in a substantial increase in the logical error rate, we instead adopt pruning and retraining~\cite{chen2023otov2-structured-pruning,lee2019signal-unstructured-pruning,liang2024automatic-structured-pruning}. Retraining is performed after initializing the remaining layers with the unpruned model weights, as this approach was observed in initial tests to converge more quickly to the original logical error rate.

\begin{table}[]
\caption{Layer-wise sparsity of GraphConv layers across different code distances, defined as the fraction of features that are zero at least 80\% of the time after ReLU. Avoidable multiplies column is the number of potential savings in the number of multiplications due to stuck on zero output features in the case of $d=7$.}
\centering
\small
\begin{tabular}{ccc}
\hline
 & \textbf{Activation Sparsity} & \textbf{Avoidable} \\
\textbf{Layer}               & \textbf{Probability}        & \textbf{Multiplies} \\
\hline
GraphConv0 & 31\% & $100\times N$ \\
GraphConv1 & 28\% & $2{,}304\times N$ \\
GraphConv2 & 49\% & $32{,}000\times N$ \\
GraphConv3 & 54\% & $141{,}312\times N$ \\
GraphConv4 & 67\% & $351{,}232\times N$ \\
GraphConv5 & 73\% & $190{,}464\times N$ \\
GraphConv6 & 88\% & $115{,}200\times N$ \\
\hline
\end{tabular}
\label{tab:sparsity-per-layer}
\end{table}

\subsection{Latency-Bounded Input Graph Filtering}
\label{subsec:input-graph-optimisation}
After the number of layers, the second most impactful parameter on the total number of operations is the maximum supported input graph size, as the overall computational cost scales linearly with the number of nodes. This is because each graph convolution layer applies the same operations to every node, as previously shown in Equation~\ref{eq:gcn} and Table~\ref{tab:comparison-gnn}. \par
By evaluating if it is necessary to consider the worst-case scenario in terms of maximum nodes supported, and possibly optimizing it, it is possible to significantly reduce the total number of iterations. Specifically, rare, more complex graphs that have a negligible impact on the final logical error rate can be discarded without processing. This eliminates the need to provision extra resources for such uncommon cases and reduces the maximum latency that must be considered. We evaluate the input-graph statistics and the effect of discarding these graphs in Section~\ref{subsec:eval-input-graph}.

\subsection{Post-Training Quantization}%0.5
\label{subsec:quantisation}
The bitwidth of the weights and each layer's output features directly affects resource utilization and computation latency. The allocation of the available Block RAMs~(BRAMs), Digital Signal Processors~(DSPs), Flip-flops~(FFs), and Look-up tables~(LUTs) can be significantly optimized by evaluating and optimizing the bitwidth of the different parts of the system. Since the number of required operations and stored parameters remains significant even after layer pruning and input graph optimization, further system-level optimization is necessary.
To identify the optimal bitwidth for the different parts of the system, we conducted a more detailed design space exploration, with a post-training quantization~\cite{gholami2021surveyquantizatio} of output features, weights, and biases of the GNN. \par
Starting from the software model, which uses floating-point computations, we move to fixed-point to reduce conversion overhead and computational complexity. We then study the effect of aggressive quantization on logical error rate and identify the best trade-off between minimizing logical error rate and meeting resource-utilization and latency constraints.
We first evaluate output features, weight, and bias quantization independently by applying fixed-point formats to one component at a time while keeping the others in single-precision floating point. We then combine the optimal configurations and evaluate the joint quantization of all three components. Although the quantization effects across components are not independent, combining the optimal configurations of each component provides a practical and effective starting point for the joint evaluation, significantly reducing the design space and avoiding a combinatorial explosion.

\subsection{Hardware Decoder Architectural Optimizations}%0.5
\label{subsec:hardware-reuse}
%After the system-level optimizations and considerations, we move towards the hardware-level optimization. 
Figure~\ref{fig:reuse-arch} shows our proposed three-stage pipelined architecture, and the function of each stage is detailed below.
% , as reported in Table~\ref{tab:cycles-per-layer}. 
\subsubsection{Input selection}
In this stage, we select the node(s) that will be computed in the following stages. In the first cycle of each new layer, we updated the input nodes with the values of the \texttt{output node register}. For the \texttt{GraphConv} layers, the computation strategy is selected based on the number of multiplications required per node: we process either multiple nodes in parallel, a single node together with the precomputed aggregated neighbors, or only one of the two. In the latter case, the computation may be further split across multiple cycles by partitioning the computations of the output features when required. The aggregated neighbors, defined as the product of the edge weights and the neighbors of each node, need to be computed before each node execution. \par
The \texttt{Global Mean-Pooling (GMP)} layer is implemented by first accumulating each feature element across all nodes and then normalizing the result by the number of nodes. Accordingly, we introduce a dedicated adder tree in the first pipeline stage. \par
For the \texttt{Dense} layers, the graph was reduced to a single node; however, the computation of the output features may still be split across multiple cycles, based on the required number of multiplications.

\subsubsection{Multipliers stage}
In this stage, since multiplication is the dominant operation in each layer, we instantiate a multiplier array sized to the available DSP resources. Mapping multiplications to DSPs is essential, as DSP blocks are FPGA resources specifically optimized for arithmetic operations such as multiplication. If the number of required multipliers exceeds the available DSP budget, the excess multiplications are synthesized in LUTs, causing a rapid increase in LUT utilization that can easily become the limiting factor for fitting the design.\par
If the number of available DSPs is less than the maximum number of computations required for some layers, we apply folding~\cite{parhi2002synthesis-folding}.  By partitioning the output columns into groups and computing one group per cycle, we ensure that the number of multiplications executed per cycle matches the available number of DSPs. \par
% In the \texttt{graph\_conv} layers, the multipliers compute the product between the neighbor nodes and the edge weights, the current node and the node feature weights matrix, or the aggregated neighbors and the node neighbor feature matrix. 
% In the \texttt{GMP} layer, the accumulated features are normalized by multiplying each feature element by a normalization factor stored in the weights BRAMs, which depends on the number of nodes. 
% In the case of the \texttt{Dense} layer, we compute the product between the input features and the weights matrix, or part of it, based on the size of the layer. The first \texttt{Dense} layer \texttt{Multipliers stage} cycle uses the \texttt{GMP} layer feed-forwarded output from the following stage. Two of the adder tree stages used for the reduction of the output of the \texttt{graph\_conv} and \texttt{Dense} layers are computed already in this stage, as the final adder tree critical path, the longest critical path in the architecture, has been divided between the two stages to reduce the maximum latency.

\subsubsection{Adder tree stage}
This stage contains the remaining stages of the adder tree. As described above, \texttt{GraphConv} is executed in two phases: the self contribution, including bias addition, is computed and stored first, followed by computation and accumulation of the neighbor contribution. In \texttt{Dense} layers, the bias is always added during the final accumulation. As several layers have a compatible output shape, we optimize the number of adder trees required, based on the final pruned GNN architecture.
%\par 
The GNN final output requires a final sigmoid activation function to obtain the result.

\begin{figure*}[ht]
\centering
\includegraphics[width=1\textwidth]{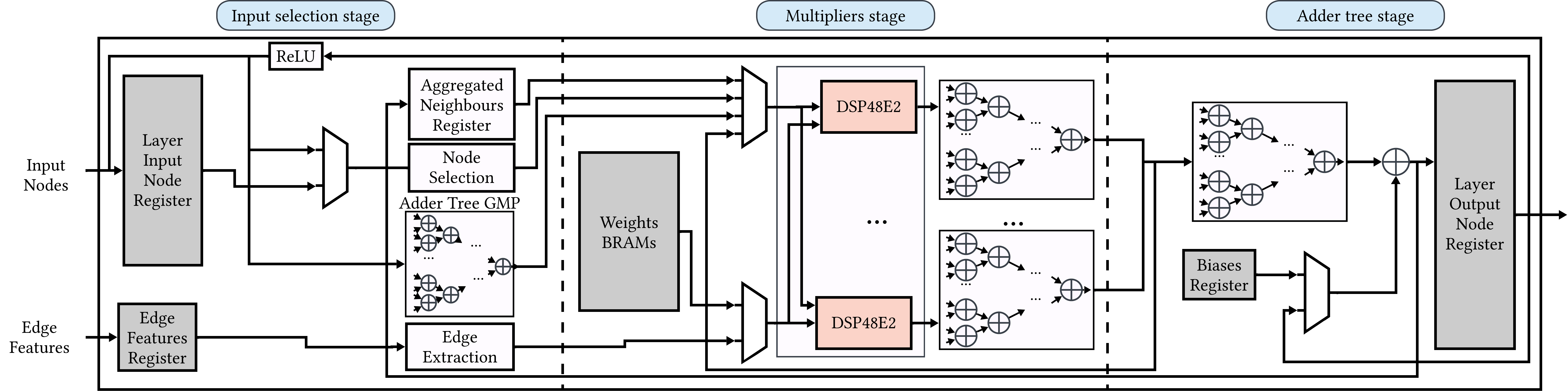}
% \Description[The figure presents the hardware microarchitecture of the GNN inference pipeline, showing the input selection, multiplication, and adder-tree stages used to implement a graph convolution layer on FPGA.]{The figure details the FPGA hardware architecture used to implement GNN inference for graph convolution. Input node features and edge features are first stored in dedicated registers and processed in the input selection stage, where neighbor features are aggregated, and partial sums are formed using an adder tree. Aggregated features are combined with weights stored in BRAMs and streamed to a bank of DSP blocks in the multipliers stage, which performs the core multiply–accumulate operations. The resulting partial products are then reduced in the adder-tree stage, where bias terms are added, and final accumulations are completed. The computed outputs are written to the layer output register, enabling pipelined execution across layers while meeting strict latency and resource constraints.}
\caption{Three pipeline stages architecture of the GNN.}
\label{fig:reuse-arch}
\end{figure*}
\subsubsection{Other design choices and optimizations}
\label{subsec:other-optimisations}
Starting from the initial architecture, we describe the multiple optimized design choices that aim at either reducing the computation latency or resource utilization.
Importantly, this level of customization does not restrict our design to a specific FPGA model. The architecture is fully parametrized with respect to the available DSP resources, enabling portability across different platforms. 

\paragraph{Interleaved edge computation}
Computing the \texttt{GraphConv} layers requires first computing the product between the edge weights and the selected node neighbors. As the number of multiplications required for this step is non-negligible, we need to use the multiplication kernel to allow for maximum parallelism. To save some cycles, we can first compute multiple aggregated neighbors at the same time, and secondly, we can move the computation of the new nodes to two cycles before the end of the current batch of nodes, as due to the pipelining, the new aggregated neighbors' value will not be saved until the new node computation starts.

\paragraph{Weight and Bias Storage}
Weights are stored in the available BRAMs. To maximize parallel computation, each cycle must supply a number of weights equal to the maximum number of concurrent multiplications. Multiple weights are packed into each memory address, and the weights stored in the BRAMs are organized such that, in each multiplication cycle, all required weights can be accessed by reading the same address across multiple parallel BRAMs.

\paragraph{Global Mean-Pooling Division}
To eliminate expensive runtime division in the \texttt{GMP} operation, we replace the division with precomputed normalization factors stored in the BRAMs alongside the weights. Because much of the BRAMs address space is unused, these scaling factors are placed at a dedicated base address, and the input graph count is used as an address offset to select the appropriate factor. Although this does not affect the latency directly, it reduces the number of multiplexers needed to control the input of the DSPs, allowing the design to meet the stringent LUT budget. \par
In summary, we presented in this section our proposed three-stage FPGA architecture, and the associated design choices targeted at optimizing the resource utilization to achieve the low-latency constraints.
% \paragraph{Adder tree stage}
% In the adder tree stage the different layers require a different configuration of adder trees to reduce the multiplication output. By doing this we are able to reduce 100k LUT per adder tree.
% \textcolor{red}{few lines about how although the output of the layers has different shapes we can basically use one single adder tree plus some extra adders to fit all of them. This reduces further the utilization}

\section{Evaluation}%2pg
\label{sec:eval}
In this section, we assess the impact of each optimization described in Section~\ref{sec:hardware-implementation} on the FPGA-based GNN complexity, decoder inference latency, and overall logical error rate. We report the results for both \singlegraph{} and \multgraph{} decoders for code distance $d=7$. 
% , which is the most sensitive to design variations and the most computationally demanding due to its larger graph size; moreover, its higher code distance results in a lower logical error rate, making it the most significant case. 
% Therefore, we will focus on the evaluation and design for code distance $d=7$.
\subsection{Evaluation Setup}%0.5
% To compare the logical error rate across different decoding approaches, we adopt a consistent evaluation workflow. All methods are tested 
% %on the same dataset generated using 
%  generating a dataset with Stim~\cite{gidney2021stim}, with a physical error rate of $p=10^{-3}$, code distance $d=3$, and circuit-level noise.
%  % a total of $n = 10^6$ samples. 
The evaluation of the reported logical error rates has been done with simulated syndrome measurements generated by Stim~\cite{gidney2021stim}, with circuit-level noise model, physical error rate of $p=10^{-3}$, and input graph sample size of $10^{8}$,
% , and sample size of $10^{7}$ for d=3, 5, 
coherently with the evaluation of the GNN~\cite{lange2023data} we selected for our work.\par
% The evaluation of the software performance in terms of latency and logical error rate was done the original work PyTorch~\cite{paszke2019pytorch} model and weights, made publicly available by the authors. 
% \par
The GNN and MWPM~\cite{higgott2022pymatching} software decoder inference time shown in Table~\ref{tab:accuracy-and-latency-comparison} have been computed on a system equipped with an Intel i9-12900K CPU and an NVIDIA RTX A4000 GPU, while the GNN training was done on a cluster with nodes equipped with Intel(R) Xeon(R) Gold 6338 CPU and NVIDIA Tesla A40 GPU.\par
 The FPGA-based decoder is designed in VHDL, and is simulated and synthesized using Vivado 2023.1, targeting the Xilinx Alveo U250 FPGA (device model: xcu250-figd2104-2L-e), a commercially available, high-end accelerator card, whose total resources are reported in Table~\ref{tab:resources}.
% We selected this toolchain version for compatibility with GNNBuilder, and 
We chose the U250 as it is representative of modern FPGAs in terms of logic capacity, memory bandwidth, and on-chip resources.

\begin{table}[tb]
% \small
    \caption{Alveo U250 resources summary.}
    \centering
    \begin{tabular}{cccc}
    \hline
       \textbf{LUT} & \textbf{FF} & \textbf{BRAM} & \textbf{DSP}\\
    \hline
        $1{,}728{,}000$ & 3{,}456{,}000 & 2688 & 12{,}288\\
    \hline
    \end{tabular}
    \label{tab:resources}
\end{table}
% \subsection{Bottleneck characterization}
% \label{subsec:bottleneck}
% If we consider the selected FPGA, Alveo U280, we can easily define the upper bound for the possible number of computations that we can execute in under a microsecond~(the threshold time).
% Assuming a clock frequency of \(f_{\mathrm{clk}} = 300\,\mathrm{MHz}\) and full utilization of the available DSP resources, the ideal throughput is up to \(2.7 \times 10^{6}\) multiplications per decoding cycle. 
% Even in these settings, this value is far from the number of computations required by the original GNN, which is in the order of magnitude of $10^{8}$.

\subsection{Hardware-Guided Pruning and Retraining}
\begin{figure}[tb]
    \centering
    \includegraphics[width=0.9\linewidth]{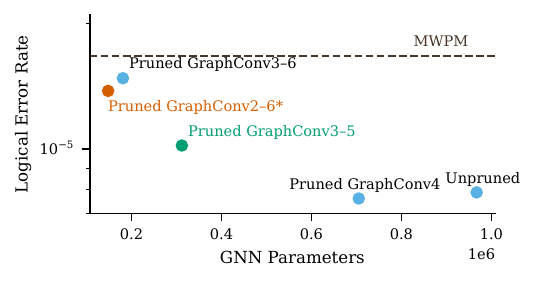}
    % \includegraphics[width=0.9\linewidth]{img/placeholder.jpg}
    % \Description[The figure shows the trade-off between GNN model size and logical error rate, highlighting how pruning reduces the number of parameters while maintaining or improving decoding performance relative to MWPM.]{The figure plots the logical error rate as a function of the number of GNN parameters for several pruned and unpruned model variants. Each point corresponds to a different GNN configuration, with progressively more aggressive pruning reducing the model size. The dashed horizontal line indicates the logical error rate achieved by the MWPM decoder, serving as a baseline. The results show that appropriately pruned models achieve logical error rates comparable to or lower than MWPM with substantially fewer parameters, illustrating an effective accuracy–complexity trade-off and motivating pruning as a key enabler for low-latency, resource-efficient decoding.}
    \caption{Comparison of logical error rates and number of different pruned and retrained models for code distance $d=7$, highlighting the two selected configurations. *Note: Layer \texttt{GraphConv2} is not entirely pruned, but half of its outputs are computed, and has been retrained for $5\times$ more epochs compared to the other models.}
    \label{fig:ler-pruned-expl}
\end{figure}

Starting from the original GNN, we evaluate progressively more aggressive pruning by considering configurations with an increasing number of removed layers, shown in Figure~\ref{fig:ler-pruned-expl}. \par
The number of training epochs (training cycles) differs across configurations. The three most complex pruned configurations have been retrained for $1{,}000$ epochs, as they already showed sufficient recovery in logical error rate, and additional training is expected to provide only marginal gains. The smallest configuration instead has been retrained for $5{,}000$ epochs because, after the initial retraining phase, its logical error rate did not recover any of the advantage over the baseline, and we therefore explored whether longer retraining could recover additional accuracy. This extra effort is motivated by prior hardware characterization of layers of different sizes, which provided early estimates of FPGA cost and latency and showed how this last configuration was the most promising for the \texttt{\singlegraph{}
} decoder. \par
% This extra effort is also supported by an 
Guided by these same hardware estimates and by the prior configurations evaluation, we select the two final models, the GNN \texttt{Pruned} \texttt{GraphConv2-6*} for the \singlegraph{} decoder and the GNN \texttt{Pruned GraphConv3-5} for the \multgraph{} decoder. 
The former is obtained from the initial GNN by pruning layers
% initial evaluation of the possible configurations and of the available resources, also based on the other optimizations, for the \singlegraph{} decoding scheme we picked the model with the layers 
\texttt{GraphConv3} through \texttt{GraphConv6}, and additionally partially pruning \texttt{GraphConv2}. Specifically, we compute only the $50\%$ of \texttt{GraphConv2} output features with the lowest activation sparsity, while the skipped feature positions are set to zero. This preserves the 256-dimensional representation expected by the subsequent GMP and dense layers while halving the multiplication count of \texttt{GraphConv2}. \par
The \multgraph{} decoder model instead is obtained by pruning the layers \texttt{GraphConv3}, \texttt{GraphConv4}, and \texttt{GraphConv5}.
% We define this configuration as \texttt{gnn-\singlegraph{}} decoder. Retraining was done until the model showed an improvement smaller than $1\%$ in logical error rate after $100$ training epochs. 
These two selected configurations maintain an improvement of logical error rate over MWPM of respectively $21\%$ and $43\%$, while greatly reducing the number of parameters and required multiplications compared to the unpruned GNN of $89\%$ and $78\%$. This translates into an estimated speed-up of $6.5\times$ and $3.1\times$. The per-layer computations of both models are reported in detail in Table~\ref{tab:computations-per-layer}.
% For the \multgraph{} decoding scheme, we picked instead the model with the layers \texttt{GraphConv4}, \texttt{GraphConv5}, \texttt{GraphConv3} pruned. We define this configuration as \texttt{gnn-\multgraph-decoding}.
\begin{table}[tb]
\caption{Number of multiplications required for each layer of the two GNNs. GraphConv layers are reported as a function of the number of nodes $N$ of the input graph.}
\centering
\begin{tabular}{ccc}
\hline
 & \multicolumn{2}{c}{\textbf{Multiplications}} \\
\textbf{Layer} & \singlegraph{} & \multgraph{} \\
\hline
GraphConv0 & $320\times N$     & $320\times N$ \\
GraphConv1 & $8{,}192\times N$    & $8{,}192\times N$ \\
GraphConv2 & $32{,}768\times N$   & $65{,}536\times N$ \\
GraphConv6 & - & $131{,}072\times N$ \\
GMP        & $256$             & $256$ \\
Dense0     & $65{,}536$           & $65{,}536$ \\
Dense1     & $32{,}768$           & $32{,}768$ \\
Dense2     & $8{,}192$            & $8{,}192$ \\
DenseOut   & $64$              & $64$ \\
\hline
\end{tabular}
\label{tab:computations-per-layer}
\end{table}
\par
We report in Table~\ref{tab:pat-improvement} the reduction of parameters and the logical error rate comparison previously shown in Figure~\ref{fig:ler-pruned-expl} with the MWPM for the two newly selected configurations.
% from $1.4\times10^7$ to $1.5\times10^6$, while still maintaining for  $d=3$, $5$, and $7$ respectively a $22\%$, $51\%$, and \imprgnnpruned{} advantage in terms of logical error rate over the MWPM decoder.

\begin{table}[tb]
% \small
    \caption{Pruning and retraining configurations and logical-error rate~(LER) for the three decoders and the baseline MWPM decoder.}
    \centering
    \begin{tabular}{cccc}
    \hline
       \textbf{Model} & \textbf{Parameters} & \textbf{LER}\\
    \hline
        MWPM & - &  \lermwpm \\
        GNN unpruned & $9.7\times10^5$ & \lergnnsw \\
    
        GNN \texttt{\singlegraph{}} &  $1.5\times10^5$ & \lergnnpruned\\
    
        GNN \texttt{\multgraph{}} & $3.1\times10^5$ & \lerparallel\\
    \hline
    \end{tabular}
    \label{tab:pat-improvement}
\end{table}

\subsection{Latency-Bounded Input Graph Filtering}
\label{subsec:eval-input-graph}
\begin{figure}
\centering
\includegraphics[width=0.9\linewidth]{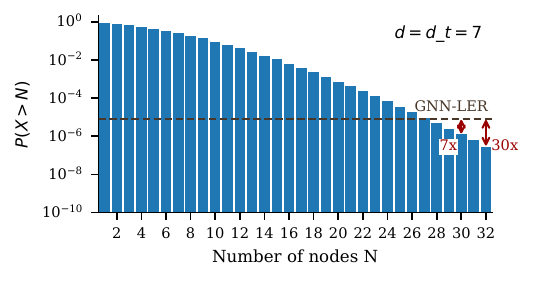}
% \includegraphics[width=0.9\linewidth]{img/placeholder.jpg}
% \Description[The figure shows the probability distribution of the number of nodes in the GNN input graph for a surface code with distance d=dt=7.]{The figure plots the probability P(n) that an input graph contains n nodes for a surface code of distance d=dt=7. The distribution is skewed toward smaller graphs, with progressively lower probability for larger node counts. The dashed horizontal line indicates the logical error rate achieved by the MWPM decoder, used as a reference baseline. The highlighted region marks the operating point of the GNN decoder, illustrating the range of graph sizes it must handle in practice.}
\caption{{Tail probabilities of the GNN input graph node count for code distance $d=7$, where $P(X > N)$ is the probability that an input graph exceeds $N$ nodes. The figure highlights the points at $N=30$ and $N=32$, corresponding to the maximum supported node counts for the \texttt{\singlegraph{}} decoder and the \texttt{\multgraph{}} decoder, respectively, and compares their tail probabilities against the logical error rate of the unpruned GNN, as these probabilities correlate with the error introduced by discarding larger graphs.}
}

\label{fig:node-distribution}
\end{figure}

To evaluate the effect of limiting the maximum acceptable input graph size on the logical error rate, we analyze the probability that a graph contains more nodes than the selected threshold, as shown in Figure~\ref{fig:node-distribution}. We notice that the tail probability decreases rapidly with $N$ and, for $N \geq 28$, falls well below the target logical error rate. This is significant because the tail probability also approximates the error contribution introduced by discarding larger graphs, which quickly becomes negligible relative to the target logical error rate.\par
% As shown in Section~\ref{subsec:eval-input-graph}, the supported input graph size can be limited to $8$, $15$, and $30$ nodes for code distances d=$3$, $5$, $7$, respectively.
% To evaluate the input graph to our system, we generate the statistics of $10^{8}$ input graphs, according to the evaluation setup.
% surface code graphs generated using Stim~\cite{gidney2021stim}, with a probability of physical error equal to $10^{-3}$ for the code distances $d=3,5,7$.The number of graphs is chosen following Lange et al.~\cite{lange2023data}, according to the target logical error rate. 
% As shown in Figure~\ref{fig:node-distribution}
Starting from this evaluation, we pick two design points, one for the \texttt{\singlegraph{}} decoder and one for the \texttt{\multgraph{}} decoder. For the former, we pick $N=30$, since the associated tail probability is nearly one order of magnitude below the target logical error rate, while larger graph sizes were found to incur excessive total latency during subsequent explorations. For the latter, we instead choose $N=32$, since the looser latency constraint allows support for larger graphs and reduces the tail probability to more than $30\times$ below the target logical error rate. Further increasing $N$ is instead limited by resource utilization, which becomes the dominant constraint once the latency target is relaxed.\par
% which is more than an order of magnitude lower than the unpruned GNN logical error rate, \lergnnsw. The number of graphs generated by running the $10^8$ Stim simulations is instead zero. While for the \texttt{\singlegraph{}} decoder configuration we consider $30$ as the maximum number of supported nodes due to the latency constraints, we can consider for the \texttt{\multgraph}{} decoder as the maximum number of supported nodes $32$. In the first case, this implies that the probability of encountering an error due to an unsupported, overly large graph is lower than the target logical error rate in the first case, and statistically improbable in the second case. 
We can therefore limit the input size of the graphs and consider a better-than-worst-case scenario in terms of execution time and required hardware. Input graphs larger than the supported size are discarded without being processed by the decoder, and a default no-error outcome is assumed. Any errors introduced by this choice are included in the reported logical error rate. This default is selected because our evaluation shows that no-error outcomes are statistically more likely than error outcomes on average. \par   
This choice significantly reduces the hardware resources required to store and process large graph instances by $\approx80\%$ for both configurations. For the \texttt{\singlegraph{}} decoder configuration, this results in a comparable decrease in maximum latency. However, it also leads to a $2\%$ increase in logical error rate relative to MWPM, reducing the resulting improvement over MWPM to $19\%$. In contrast, for the \texttt{\multgraph{}} decoder configuration, graphs exceeding this threshold occur so infrequently that the corresponding increase in logical error rate is negligible and remains within the rounding error of the initial improvement value of $41\%$.

\subsection{Post-Training Quantization}
\label{subsec:eval-ptq}
\begin{figure}[t]
    \centering
    \includegraphics[width=0.9\linewidth]{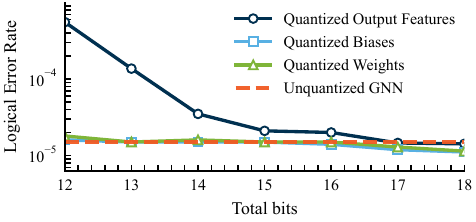}
    % \Description[The figure shows the impact of quantizing different GNN components on the logical error rate as a function of total bit precision.]{The figure plots the logical error rate versus total bitwidth for quantized GNN components. Quantizing output features has the largest impact at low precision, while quantizing weights and biases shows minimal degradation across the evaluated range. The dashed line indicates the logical error rate of the unquantized GNN, serving as a performance reference. These results identify the minimum precision at which quantization does not affect decoding performance.}
    \caption{Effect of quantizing only weights, layers output features, or biases on the logical error rate for $d=7$, for GNN \texttt{\singlegraph} decoder.}
    \label{fig:quant-effect}
\end{figure}

% \textcolor{green}{\lipsum[18-22]}
The results of the post-training quantization independently applied to weight, bias, and output features of the \texttt{\singlegraph{}} decoder are shown in Figure~\ref{fig:quant-effect}. While bias and weights can still achieve the same logical error rate as the unquantized model, the quantization of the output features leads to a loss. After a design space exploration of the fully quantized model based on the previous analysis, we observed that the best quantization settings that optimize resource usage, reduce resource latency, and minimize logical error rate loss are the following:
\begin{itemize}
    \item Weights quantized to 14-bit fixed-point with 4 integer bits and 10 fractional bits,
    \item Output features quantized to 17-bit fixed-point with 12 integer bits and 5 fractional bits,
    \item Biases quantized to 5-bit fixed-point with 1 integer bit and 4 fractional bits.
\end{itemize}
We keep biases at the same quantization as the data, as more aggressive quantization provides little benefit in terms of resource usage. This is due to the limited number of bias values and the requirement to extend them to the data precision to be added. Accumulation is done with 27 bit fixed point, with 12 integer bits and 15 fractional bits. The output features are rounded after accumulation.\par
For the \texttt{\multgraph{}} configuration, average latency remains an important consideration, but resource utilization becomes the dominant bottleneck, still requiring a custom fixed-point quantization scheme to meet the target. Based on an analysis analogous to the previous one, we choose the same quantization scheme for weights and biases as for the \texttt{\singlegraph{}} decoder, while the data representation uses 18 integer bits and 5 fractional bits. The accumulation is done with 28 bit fixed point, with 13 integer bits and 15 fractional bits, and rounding after accumulation.\par
This optimization improves resource utilization mainly by reducing the DSP cost of each multiplication from four units to one. On the Alveo U250, 32-bit operands require four DSPs per multiply, while multiplications whose total operand width does not exceed 48 bits can be implemented with a single DSP; beyond this threshold, the compiler maps a single multiplication onto multiple DSPs. As a result, mapping one multiplication per DSP allows a $4\times$ increase in parallel multiplications and an approximately equivalent speed-up in terms of total latency. It also leads to a decrease in the maximum latency of the adder tree. \par
To achieve this, we have a trade-off in the logical error rate of both models, which, after quantization, have an advantage of  \imprgnnprunedquantizednodes{} for the \texttt{\singlegraph{}} decoder and of $40\%$ in the case of the \texttt{\multgraph{}} decoder.
% This configuration allowed us to optimise DSP utilisation, as detailed in Table~\ref{tab:dsp-intervals}, and achieve a $4\times$ increase in the number of parallel multiplications.

\subsection{Hardware Optimizations}
%As discussed in earlier sections, we apply several optimization steps to optimally use the resources to decrease the latency.
The architecture shown in Figure~\ref{fig:reuse-arch} is used for both decoders, with differences only in the bitwidths adopted, as described in Section~\ref{subsec:eval-ptq}, and in the maximum supported input-graph size, as discussed in Section~\ref{subsec:eval-input-graph}.
\paragraph{Adder trees}
Because the output adder tree has a power-of-two structure in every layer except \texttt{GraphConv0}, we can improve resource utilization through hardware reuse.
% \begin{itemize}
%     \item \texttt{GraphConv2}, \texttt{dense\_2} can be computed by using 4 32-input adder trees, by adding two extra final stages
%     \item \texttt{dense\_0}, \texttt {dense\_1} can be computed by using 8 32-input adder trees, by adding three extra final stages
%     \item \texttt{dense\_out} can be computed by using 2 32-input adder trees, by adding one extra final stage
%     \item The output of the multiplication of the neighbors by the weight edges can be computed directly by feeding zeroes to the unused inputs, as it requires $150$, $256$, and $256$ 30-input adder trees, for the \texttt{GraphConv0}, \texttt{GraphConv1}, and \texttt{GraphConv2} layers.
% \end{itemize}
Although this does not affect the overall latency, it strongly affects the overall LUT utilization.

% \begin{table}[t]
% \centering
% \caption{Adder trees configuration for computation of nodes and aggregated neighbors.}
% \label{tab:adder-tree-config}
% \small
% \begin{tabular}{lcccc}
% \hline
% \multicolumn{3}{c}{\textbf{Node}} & \multicolumn{2}{c}{\textbf{Aggregate neighbors}} \\
% \hline
% \textbf{Layer} & \textbf{Adder Trees} & \textbf{Inputs} & \textbf{Adder Trees} & \textbf{Inputs} \\
% \hline
% GraphConv\textsubscript{0}   & 1600 & 5   &  5   & 30 \\
% GraphConv\textsubscript{1}   & 256  & 32  &  32  & 30 \\
% GraphConv\textsubscript{2}   & 64   & 128 &  128 & 30 \\
% Dense\textsubscript{0} & 32   & 256 &      &    \\
% Dense\textsubscript{1} & 32   & 256 &      &    \\
% Dense\textsubscript{2} & 64   & 128 &      &    \\
% Dense\textsubscript{out} & 1    & 64  &      &    \\
% \hline
% \end{tabular}
% \end{table}

\begin{figure}[h]
    \centering
    \includegraphics[width=0.9\linewidth]{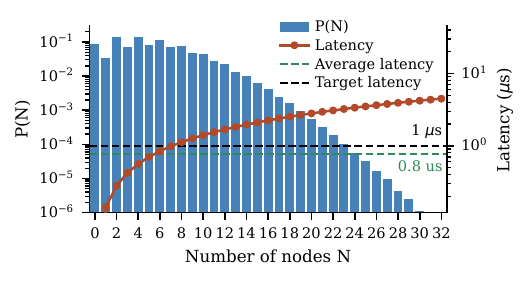}
    \caption{Latency as a function of the number of input graph nodes $N$ for the \texttt{\multgraph} decoding configuration. While latency grows with the increasing number of nodes, high-node-count syndrome cases occur infrequently, leading to an average latency of \qty{0.8}{\micro\second}, which remains below the target \qty{1}{\micro\second}.}
    \label{fig:probability-latency-distribution}
\end{figure}

\paragraph{BRAMs organization}
Based on the quantization results, we pack five 14-bit weights into each memory address. Supporting 8,192 parallel multiplications, therefore, requires reading an equal number of weights per cycle, which corresponds to accessing 1,639 BRAMs out of the 2,688 available. Although this represents a high fraction of the available BRAMs, the utilization of each BRAM’s address space remains low, leaving sufficient capacity to store the GMP normalization factors as discussed earlier. Bias values, by contrast, are stored directly in registers due to their limited count. We report the \texttt{\singlegraph{}} decoder final cycle count for the worst case with $N=30$ in Table~\ref{tab:cycles}, and the \texttt{\multgraph{}} decoder average latency in Figure~\ref{fig:probability-latency-distribution}.

\begin{table}[]
\caption{Number of clock cycles and latency required for each layer in case of $d=7$, $N=30$, and $t_{clk}=$\qty{4.8}{\nano\second} for the \texttt{\singlegraph{}} decoder configuration.}
\centering
% \small
\begin{tabular}{ccc}
\hline
\textbf{Layer} & \textbf{Clock Cycles} & \textbf{Latency~[ns]}  \\
\hline
GraphConv0 & 7 & 33.6\\
GraphConv1 & 38 &182.4\\
GraphConv2 & 137 & 657.6\\
GMP & 2 & 9.6\\
Dense0 & 10 &48\\
Dense1 & 6 &28.8\\
Dense2 & 3 & 14.4\\
DenseOut & 3 & 14.4\\
\hline
\textbf{Total} & 206 & 988.8 \\
\hline
\end{tabular}

\label{tab:cycles}
\end{table}

\subsection{Summary}
\begin{figure}
    \centering
    \includegraphics[width=0.9\linewidth]{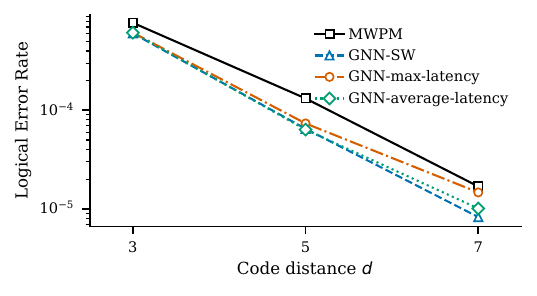}
    \caption{Logical error rate versus code distance comparing MWPM, original GNN~(GNN-SW), and the two optimized GNN decoders. Both optimized GNNs outperform MWPM across the evaluated distances.}
    \label{fig:d_vs_ler}
\end{figure}

\begin{figure}[t]
    \centering
    \includegraphics[width=0.9\linewidth]{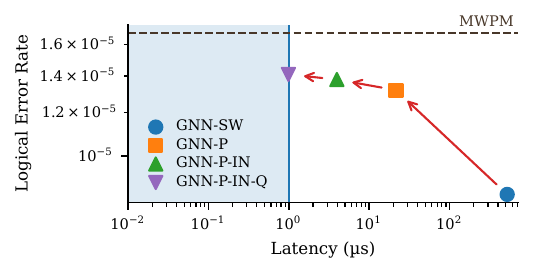}
    % \Description[The figure compares logical error rate versus decoding latency for different GNN decoder implementations, highlighting the accuracy–latency trade-offs relative to MWPM.]{The figure plots logical error rate as a function of decoding latency for several GNN-based decoder variants. Different implementations explore trade-offs between inference speed and decoding accuracy, spanning software, pruned, and quantized designs. The dashed line denotes the MWPM baseline, illustrating how GNN decoders achieve comparable or improved error rates under strict latency constraints.}
    \caption{Comparison of logical error rate and latency across different optimizations of \texttt{\singlegraph{}} decoder: original GNN~(GNN-SW), pruned GNN~(GNN-P), pruned GNN with reduced maximum input nodes~(GNN-P-IN), and pruned GNN with reduced maximum input nodes and quantization~(GNN-P-IN-Q). GNN-P and GNN-IN latencies are estimated from operation counts.}
    \label{fig:latency-vs-ler}
\end{figure}

Table~\ref{tab:synth} reports the synthesis results for both configurations, including resource utilization and the achieved clock period. Figure~\ref{fig:d_vs_ler} shows the comparison between the LER of the two models across code distances up to $d=7$. Figure~\ref{fig:latency-vs-ler} highlights the optimization steps and their effect on the \texttt{\singlegraph{}} configuration. The final design point includes the final hardware-level optimization that reduces execution to 206 cycles, allowing the decoder to reach the sub-microsecond threshold with a latency of \qty{988.8}{\nano\second} and a logical error rate of \lergnnprunedquantizednodes{} in the context of \singlegraph{} decoding. In the context of \texttt{\multgraph{}} decoding we achieve instead an average latency of \qty{846}{\nano\second}, with a logical error rate of $1.01\times10^{-5}$. We are therefore able to decode in time under the two settings, achieving an improvement over the MWPM logical error rate of \imprgnnprunedquantizednodes{} for the \texttt{\singlegraph{}} decoder and $40\%$ for the \texttt{\multgraph{}} decoder.

% We also report the number of cycles per node for each layer in Table~\ref{tab:cycles-per-layer}. 

\begin{table}[t]
\small
    \caption{Synthesis results for \texttt{\singlegraph{}} decoder~(a) and \texttt{\multgraph} decoder~(b) on Alveo U250 FPGA.}
    \centering
    \begin{tabular}{cccccc}
    \hline
       \textbf{Model} & \textbf{LUT} & \textbf{FF} & \textbf{BRAM} & \textbf{DSP} & \textbf{$t_{clk}$}    \\
    \hline
        (A) & $72.96\%$ & $11.50\%$ & $60.96\%$ & $66.67\%$ & \qty{4.8}{ns} \\
        (B) & $86.07\%$ & $16.45\%$ & $60.96\%$ & $66.67\%$ & \qty{5}{ns} \\
    \hline
    \end{tabular}
    \label{tab:synth}
\end{table}


\subsection{Scalability}
Further improvements to the decoder can be achieved either by accommodating a less aggressively pruned model or more complex GNN models, thereby reducing the logical error rate, or by supporting larger code distances. Both directions require an increase in hardware resources. In the former case, additional memory capacity and greater parallelism are needed to sustain the larger model. In the latter, scaling to higher code distances primarily demands increased parallelism to offset the additional cycles required to process the larger number of nodes.
Although the selected FPGA is close to full resource utilization, current and future FPGAs, such as the Xilinx Versal series, provide additional DSP, LUT, and BRAM resources that can be exploited in several ways. Increased compute and memory capacity enables greater parallelism, reducing per-layer latency and allowing less aggressive pruning, which can further improve the logical error rate relative to MWPM. Additional resources can also support scaling to larger code distances~\cite{lange2023data}, where worst-case error patterns increase the number of input graph nodes and operations per layer. 

% neu2024dynamically-fpga-gnn1,

\section{Related Work}
A growing body of work has studied GNN acceleration in hardware~\cite{kim2025eod-gnn-asic,kim2025omega-gnn-acceleration,zhou2022mode-gnn-fpga-2,zhang2022low-gnn-fpga-cpu,abi2023gnnbuilder}. However, only a limited subset is relevant to the problem of sub-microsecond single-instance inference on FPGA targets. Several of these works instead focus on GPU-based low-latency inference or broader system-level deployment challenges~\cite{kim2025eod-gnn-asic,kim2025omega-gnn-acceleration}, batched or throughput-oriented execution on FPGA or heterogeneous platforms~\cite{zhou2022mode-gnn-fpga-2,zhang2022low-gnn-fpga-cpu}, or toolflows that automatically map PyTorch models to FPGA implementations~\cite{abi2023gnnbuilder}. While these are complementary directions, they do not target the same latency, workload, and deployment constraints as our setting. The closest prior work is by Que et al.~\cite{que2024ll-llgnn}, which similarly focuses on co-design for sub-microsecond inference. However, the two settings differ substantially. Their design targets a much smaller GNN, with approximately $3{,}000$ parameters, allowing parallel execution of multiple layers and extensive sub-layer fusion. Our setting instead involves significantly larger layers that must be executed over multiple cycles, making resource-constrained scheduling and hardware-aware model reduction central to meeting the latency target. Moreover, their workload assumes fully connected input graphs, while our decoder operates on $k$-nearest-neighbor graphs with variable node counts. This difference is architecturally important, as it leads to input-dependent execution time and resource requirements, and therefore to a substantially different optimization problem.
% As shown in previous chapters we apply multiple optimisations which allow to reduce the utilisation, by first resizing the network and having 40x less parameters, and then by applying quantization and obtaining a size of by using DSPs. After further applying quantization, we can see that we have a proportional decrease in the number of bits, to reduce the number of DSPs needed for each operation. 

% \begin{table}
% \centering
% \begin{tabular}{|c|c|c|c|}
% \hline
% \textbf{Decoder} & \textbf{Data format }      & \textbf{DSPs} & \textbf{LUT}\\
% \hline
% SW & fp&  - & - \\
% \hline
% HW &  fxp 14bits &  2640 & 32\%\\

% HW &  fxp 26bits &  4688 & 60\%\\

% HW &  fxp 32bits &  9024 & 150\%\\
% \hline
% \end{tabular}
% \caption{Utilisation of DSPs and LUT comparison.}
% \label{tab:hardware-utilisation}
% \end{table}

% \section{Related Work}%0.75
% \label{sec:related-work}
% \input{src/s6-related}

\section{Conclusion}%0.25
\label{sec:conclusion}
% In this work, starting from a software-based graph neural network, we applied several hardware-oriented optimizations to be able to execute real-time quantum error correction. First, we were able to hugely decrease the size of the GNN by pruning multiple layers via pruning-aware training to significantly lower computational complexity, while maintaining a logical error rate advantage over MWPM decoders. After a first evaluation of the hardware bottlenecks, we focused on further optimization to be able to fit the GNN in the selected off-the-shelf FPGA, while still achieving the required latency. Firstly, we applied a uniform post-training quantization to reduce precision without compromising the logical error rate, achieving both lower hardware usage and speeding up the overall execution. Secondly, we designed a hardware architecture that maximizes resource utilization through folding and efficient scheduling. 
In this work, we start from a high-accuracy software-based GNN decoder and apply a sequence of hardware-driven optimizations to make real-time QEC feasible. These optimizations include hardware-guided pruning and retraining, input-graph filtering, post-training quantization, and several architecture-level choices to improve resource utilization and reduce the GNN-decoder inference latency.\par
Through this co-design methodology, we derived two optimized models that outperform MWPM in terms of logical error rate in both the \singlegraph{} and \multgraph{} decoding settings. In the \singlegraph{} setting, for $d=7$, our design achieves a worst-case latency of \latency per inference and a logical error rate of \lergnnprunedquantizednodes, corresponding to a \imprgnnprunedquantizednodes{} improvement over MWPM. In the parallel decoding setting, it achieves an average latency of \qty{846}{\nano\second}, a logical error rate of $1.01\times10^{-5}$, and a $40\%$ improvement over MWPM. \par
Together, these results show that GNN-based decoding can meet stringent real-time constraints while preserving its accuracy advantage in resource-constrained hardware environments. Moreover, while our design is hardware-aware, the proposed techniques are broadly applicable and are not specific to a particular FPGA platform.
% This flexibility allows our approach to scale with newer, more capable FPGA platforms, opening the door to even greater performance and efficiency.\par

% Looking at future further explorations, there is potential for additional improvement through the use of quantitation-aware training, which can help maintain accuracy under aggressive precision reduction, with likely better performance compared to the post-training quantitation in terms of bitwidth saving. In parallel, while the original model was pruned arbitrarily, there is the possibility to apply accuracy-aware pruning, a method already shown to work in other neural networks works. 

%I talk about scaling here, should it be moved to a different part of the paper?

\section*{Acknowledgment}
We acknowledge support from the Swedish Foundation for Strategic Research (grant number FUS21-0063).

\bibliographystyle{unsrt}
\bibliography{refs}

\end{document}